\documentclass[a4paper,12pt]{article}
\usepackage[utf8]{inputenc}
\usepackage[english]{babel}
\usepackage{graphicx}

\usepackage[onehalfspacing]{setspace}

\usepackage[headheight=1in,margin=1in]{geometry}
\usepackage{lipsum}
\usepackage{amsmath}

\newcommand\footnotenomarker[1]{%
  \begingroup
  \renewcommand\thefootnote{}\footnote{#1}%
  \addtocounter{footnote}{-1}%
  \endgroup
}

\usepackage[normalem]{ulem} 

\usepackage{dsfont}

\usepackage{amsfonts}

\usepackage[authoryear,sort]{natbib}
\usepackage{titlesec}
\titlespacing{\paragraph}{0pt}{0pt}{1em}
\titlespacing*{\section}{0pt}{0.\baselineskip}{0.2\baselineskip}
\titlespacing*{\subsection}{0pt}{0.\baselineskip}{0.2\baselineskip}
\titlespacing*{\subsubsection}{0pt}{0.\baselineskip}{0.2\baselineskip}

\usepackage{subcaption}
\usepackage{float}

\usepackage{hyperref}
\hypersetup{hidelinks}

\usepackage[bottom]{footmisc}
\usepackage{xcolor}

\definecolor{sunset}{RGB}{238,93,108}
\definecolor{sunset_deep}{RGB}{106,13,131}

\graphicspath{{figures/}}

\usepackage{float}

\usepackage{booktabs}

\usepackage{amsthm}
\newtheoremstyle{customassump}
  {}{}                         
  {\itshape}                   
  {}                           
  {\bfseries}                  
  {}                           
  {.5em}                       
  {}                           
\theoremstyle{customassump}
\newtheorem{assumption}{Assumption}
\newtheorem{theorem}{Theorem}
\newtheorem{definition}{Definition}

\title{\vspace{-1in}\singlespacing Anytime-Valid Inference in Adaptive Experiments: Covariate Adjustment and Balanced Power}

\author{Daniel Molitor\thanks{Department of Information Science, Cornell University, djm484@cornell.edu.} \and Samantha Gold\thanks{Brooks School of Public Policy, Cornell University, sg972@cornell.edu.}}
 
\date{\today}

\begin{document}

\maketitle
\vspace{-.25in}
\begin{center}
\end{center}

\begin{singlespacing}
\begin{abstract}
    Adaptive experiments such as multi-armed bandits offer efficiency gains over traditional randomized experiments but pose two major challenges: invalid inference on the Average Treatment Effect (ATE) due to adaptive sampling and low statistical power for sub-optimal treatments. We address both issues by extending the Mixture Adaptive Design framework of \citeauthor{liang_experimental_2024}. First, we propose MADCovar, a covariate-adjusted ATE estimator that is unbiased and preserves anytime-valid inference guarantees while substantially improving ATE precision. Second, we introduce MADMod, which dynamically reallocates samples to underpowered arms, enabling more balanced statistical power across treatments without sacrificing valid inference. Both methods retain MAD's core advantage of constructing asymptotic confidence sequences (CSs) that allow researchers to continuously monitor ATE estimates and stop data collection once a desired precision or significance criterion is met. Empirically, we validate both methods using simulations and real-world data. In simulations, MADCovar reduces CS width by up to 60\% relative to MAD. In a large-scale political RCT with 32,000 participants, MADCovar achieves similar precision gains. MADMod improves statistical power and inferential precision across all treatment arms, particularly for suboptimal treatments. Simulations show that MADMod sharply reduces Type II error while preserving the efficiency benefits of adaptive allocation. Together, MADCovar and MADMod make adaptive experiments more practical, reliable, and efficient for applied researchers across many domains. Our proposed methods are implemented through an open-source software package.
\end{abstract}
\end{singlespacing}

\section{Introduction}

\footnotenomarker{Author's note: The MAD and our proposed extensions are implemented via an open-source Python package, available
\hyperlink{https://pypi.org/project/pyssed/}{through PyPI} and on GitHub at \href{https://github.com/dmolitor/pyssed}{\textcolor{blue}{github.com/dmolitor/pyssed}}. Additionally, a replication package for all figures and tables is available at \href{https://github.com/dmolitor/anytime-valid-inference-in-adaptive-experiments}{\textcolor{blue}{github.com/dmolitor/anytime-valid-inference-in-adaptive-experiments}}.}

Adaptive experiments have become a key tool in modern experimental design. Adaptive experimental designs, such as multi-armed bandits (MABs), can help researchers achieve particular goals that are beyond the scope of standard randomized experiments: rapidly identifying the optimal treatment arm(s) from a larger set of treatments \citep{russo_simple_2018, gosciak_adaptive_2025}, improving statistical power and inferential precision for optimal treatments \citep{offerwestort_adaptive_2021}, and maximizing cumulative respondent welfare during the experiment \citep{thompson_likelihood_1933}. Although adaptive experiments are far more efficient for these stated goals than randomized experiments, they introduce two notable challenges for applied researchers.

The thorniest limitation is that adaptive experiments compromise valid causal inference, making standard average treatment effect (ATE) estimators unreliable. The primary inferential challenge is that the treatment assignment $W_t$ for unit $t$ depends on prior treatment assignments and outcomes $\{(W_1, Y_1),...,(W_{t-1},Y_{t-1})\}$. This violates core assumptions that standard ATE estimators like difference-in-means and inverse propensity score estimators rely on for unbiasedness or valid inference, rendering such estimators unreliable on adaptively collected data \citep{hadad_confidence_2021, dimakopoulou_online_2021}. Additionally, adaptive experiments typically concentrate sample on the optimal treatments (e.g. the treatment that maximizes respondent welfare), leaving sub-optimal treatments underpowered---a critical limitation when robust inference is needed across all treatments, not just the best-performing one.

Recent advances in adaptive experimental design address some of these challenges. One strand of work that addresses the challenge of valid inference in adaptive experiments (\citet{hadad_confidence_2021}, \citet{bibaut_post-contextual-bandit_2021}, \citet{zhang_statistical_2021}, \citet{zhang_inference_2021}) relies on reweighting schemes that restore asymptotically-valid confidence intervals. However, these methods demand that every treatment retain a positive sampling probability at each unit $t$, ruling out common adaptive designs such as Explore–Then–Commit (ETC) and Upper Confidence Bound (UCB) algorithms. Another, more recent, strand (\citet{waudby-smith_anytime-valid_2024}, \citet{ham_design-based_2023}) circumvents asymptotic normality altogether by employing anytime–valid inference, which maintains error control for arbitrary adaptive designs. The Mixture Adaptive Design (MAD), proposed by \citet{liang_experimental_2024}, adopts this approach and resolves the challenge of valid inference in adaptive experiments by enabling unbiased estimation of the ATE and providing valid, $(1-\alpha)$ confidence sequences (CSs) for all treatment arms.\footnote{$\alpha$ is a fixed significance level, e.g. $\alpha=0.05$.} Although MAD enables valid inference in adaptive experiments, its ATE estimates often suffer from excess variance, leading to wide CSs. Additionally, MAD does not directly address the statistical power imbalance inherent in adaptive sampling, as the underlying adaptive sampling strategy still concentrates samples on optimal treatments, leaving sub-optimal treatments underpowered.

In this paper we propose modifications to the MAD estimation framework that address both of these challenges: covariate-adjusted MAD can achieve significant ATE precision improvements via covariate adjustment, and power-modified MAD adaptively re-allocates sample to under-powered treatments to ensure improved statistical power for all treatments.

\section{Background}

\subsection{Anytime-valid inference}

It is often desirable, particularly in the context of online or sequential experiments, to continuously monitor parameter estimates in real-time with the goal of dynamically stopping data collection once a desired precision or significance criterion is met.

Confidence intervals (CIs) are well-known frequentist statistical tools that quantify the uncertainty around parameter estimates. Traditionally, a 95\% confidence interval computed from a fixed sample size provides a range of plausible values for the parameter of interest, guaranteeing that, under repeated sampling, approximately 95\% of such intervals will contain the true parameter. Precisely, a $(1 - \alpha)$ CI $[L, U]$ for parameter $\delta$ guarantees that for a given sample size $n$ Pr$(\delta \in [L, U]) \geq 1 - \alpha$. However, conventional CIs are only valid when $n$ is fixed beforehand; any form of continuous monitoring and dynamic stopping invalidates this inferential guarantee and inflates Type 1 error rates.

CSs are interval estimates that hold uniformly over time, \textit{i.e. anytime-valid}, meaning that their Type 1 error guarantees remain valid regardless of whether, and how often, researchers inspect the data. A $(1-\alpha)$ CS is a random sequence of intervals indexed by sample size, which has at least probability $(1-\alpha)$ of containing the true parameter simultaneously over the entire sequence. Precisely, a $(1 - \alpha)$ CS $[L_n, U_n]$ for parameter $\delta$ guarantees that, Pr$(\forall n, \delta \in [L_n, U_n]) \geq 1 - \alpha$. An important implication is that CSs allow researchers to continuously monitor parameter estimates and use these estimates to dynamically stop data collection, all while maintaining Type 1 error guarantees \citep{ramdas_1_2018}.

Next, we introduce a slightly weaker version of CSs---asymptotic CSs. Precisely, $C_n=[L_n, U_n]$ is a $(1-\alpha)$ asymptotic CS for parameter $\delta$ if there exists some (unknown) non-asymptotic $(1-\alpha)$ CS ($C_n^*$) for $\delta$ such that $C_n \overset{a.s.}{\to} C_n^*$. Asymptotic CSs require less-restrictive assumptions compared to their non-asymptotic counterparts and therefore facilitate applications in a wide array of settings. Although they sacrifice small-sample validity for flexibility and generalizability, asymptotic CSs have been shown to perform very well empirically, even in small- and moderate-sample settings (\citet{ham_design-based_2023}; \citet{waudby-smith_time-uniform_2024}; \citet{liang_experimental_2024}).

\subsection{Notation}

In this section, we formalize our problem setting and corresponding notation.

Following the notation of \citet{liang_experimental_2024}, we begin by assuming that we observe a sequence of $t$ experimental units $\{X_i,W_i,Y_i\}_{i=1}^t$ where $W_i \in \{0, ..., K-1\}$, $Y_i$, and $X_i \in \mathbb{R}^d$ are the treatment assignment ($W_i=0$ indicating assignment to the control group), observed outcome, and covariate vector for unit $i$, respectively. Although our results hold for both $K \geq 2$ (multiple treatments) as well as the batched sampling setting, for the remainder of our main discussion we will assume that we observe a single unit at a time and that we have binary treatment assignments, i.e. $W_i \in \{0, 1\}$. Throughout our analysis, we adopt the Stable Unit Treatment Value Assumption \citep{rubin_comment_1986}, which assumes no population interference and that, for each unit, there is only one version of each treatment.

Given these assumptions, each unit $i$ has a pair of potential outcomes $\{Y_i(0), Y_i(1)\}$ \citep{rubin_causal_2005}, and the observed outcome depends on the treatment assignment: \[Y_i=\mathds{1}\{W_i=1\}Y_i(1) + \mathds{1}\{W_i=0\}Y_i(0).\] The individual treatment effect (ITE) is defined as $\tau_i:=Y_i(1)-Y_i(0)$, and the ATE up to unit $t$ is \[\bar{\tau}_t:=\frac{1}{t}\sum_{i=1}^t{\tau_i}.\]

\subsection{Design-based inference}

Following \citet{liang_experimental_2024}, we adopt a \textit{design-based} perspective on causal inference, as opposed to a \textit{model-based} perspective. Under the design-based perspective, experimental units are drawn from a finite population which the researcher fully observes and the causal estimands of interest are specific to the actual sample gathered. In contrast, the model-based perspective assumes that experimental units are drawn from some (potentially infinite) super-population, and the estimands of interest are defined relative to this infinite super-population. Under the design-based perspective, the set of all covariate vectors and potential outcomes for the population can be treated as unknown but non-random quantities \citep{abadie_samplingbased_2020}. As a result, the properties of our estimator arise solely from the randomness induced by the treatment assignment policy; that is, $\{X_i, Y_i(0), Y_i(1)\}$ is non-random, so $Y_i$ is random exclusively through the randomness induced by treatment assignment $W_i$. For a deeper examination of the design-based perspective and its application to adaptive experimental design, we refer the reader to \citeauthor{liang_experimental_2024}, Section 3.1.

\section{Covariate-adjusted Mixture Adaptive Design}
\label{sec:mad_covar}

We now formalize covariate-adjusted MAD (MADCovar), which extends MAD to accomodate covariate adjustment in its ATE estimates, often resulting in significant precision gains.

As is standard in the adaptive experimental literature, we assume that treatment assignment for a given unit $t$ can depend on previously observed outcomes and treatment assignments up to $t-1$. Formally, define the history at unit $t$ as $\mathcal{H}_t := \{Y_i, W_i, X_i\}_{i=1}^t$, where $Y_i, W_i, X_i$ are previously observed outcomes, treatment assignments, and covariate vectors, and let $\mathcal{F}_t$ be the sigma-algebra containing all pairs of potential outcomes $\{Y_i(1), Y_i(0)\}_{i=1}^t$ as well as $\mathcal{H}_{t-1}$. Additionally, let $\mathcal{A}$ be an adaptive treatment assignment policy, such as Thompson sampling \citep{thompson_likelihood_1933} or the Upper Confidence Bound algorithm, and define the probability that $\mathcal{A}$ assigns unit $t$ to treatment $w$ as $p_t^\mathcal{A}(w) := \mathbb{P}_\mathcal{A}(W_t=w|\mathcal{H}_{t-1})$. Finally, let $\delta_t \in (0, 1]$ be a deterministic sequence which, for each unit $t$, determines the tradeoff between a fully random assignment policy and $\mathcal{A}$, with the requirement that $\delta_t = \omega\left(\frac{1}{t^{1/4}}\right)$.\footnote{By definition, $\delta_t = \omega\left(\frac{1}{t^{1/4}}\right)$ implies that $\frac{1}{\delta_t} = o\left(t^{1/4}\right)$. Intuitively, $\delta_t$ is a function of $t$ that converges to 0 slower than $\frac{1}{t^{1/4}}$.}

\begin{definition}{\textup{(Liang \& Bojinov; Definition 4 [Mixture Adaptive Design]).}}
\label{def:mad_probability}
For any potentially adaptive treatment assignment algorithm $\mathcal{A}$, real-valued sequence $\delta_t \in (0, 1]$, and $w \in \{0, 1\}$, the probability that the MAD will assign treatment $w$ for unit $t$ is: \[p_t^\text{MAD}(w) := \mathbb{P}_\text{MAD}\left(W_t=w | \mathcal{H}_{t-1}\right) = \delta_t \left(\frac{1}{2}\right) + (1-\delta_t)p_t^\mathcal{A}(w).\]
\end{definition}

A natural extension of Definition~\ref{def:mad_probability} to the $K \geq 2$ treatment setting with $w \in \{0, 1, ..., K-1\}$ is to define $p_t^\text{MAD}(w) := \mathbb{P}_\text{MAD}(W_t=w|\mathcal{H}_{t-1})=\delta_t\left(\frac{1}{K}\right) + (1-\delta_t)p_t^\mathcal{A}(w)$. Definition~\ref{def:mad_probability} can also be easily adapted for batched assignment, for which we refer the reader to Definition 5 (Section 4.1) of \citeauthor{liang_experimental_2024}.

\subsection{Covariate-adjusted ATE estimation}
\label{sec:ate_estimation}

We now define our ATE estimator. We adopt the following covariate-adjusted estimator of $\tau_i$ in the manner of \citet{robins_semiparametric_1995}:
\begin{equation}
    \label{eq:ate}
    \hat{\tau}_i := \hat{\mu}_1(X_i) - \hat{\mu}_0(X_i) + \frac{\mathds{1}\{W_i=1\}(Y_i - \hat{\mu}_1(X_i))}{p_i^\text{MAD}(1)} - \frac{\mathds{1}\{W_i=0\}(Y_i - \hat{\mu}_0(X_i))}{p_i^\text{MAD}(0)},
\end{equation} where $\hat{\mu}_w(X_i)$ is an arbitrary estimator of $\mathrm{E}[Y_i(w)|X=X_i,\mathcal{F}_{i-1}]$.
Under the design-based perspective, it follows that $\mathrm{E}[\hat{\tau}_i|\mathcal{F}_i]=\tau_i$, where the expectation is taken with respect to the treatment assignment mechanism, and thus $\hat{\bar{\tau}}_t:=\frac{1}{t}\sum_{i=1}^t{\hat{\tau}_i}$ is an unbiased estimator for $\bar{\tau}_t$. Additionally, we have that, for all $i$, \[\mathrm{Var}[\hat{\tau}_i|\mathcal{F}_i] \leq \sigma^2_i, \text{ where } \sigma^2_i:=\frac{(Y_i(1)-\hat{\mu}_1(X_i))^2}{p_i^\text{MAD}(1)}+\frac{(Y_i(0)-\hat{\mu}_0(X_i))^2}{p_i^\text{MAD}(0)},\] and thus, a natural unbiased estimator for $\sigma^2_i$ is \[\hat{\sigma}^2_i := \frac{\mathds{1}\{W_i=1\}(Y_i - \hat{\mu}_1(X_i))^2}{(p_i^\text{MAD}(1))^2} + \frac{\mathds{1}\{W_i=0\}(Y_i - \hat{\mu}_0(X_i))^2}{(p_i^\text{MAD}(0))^2}.\] For future definitional purposes, we will define $S_t := \sum_{i=1}^t\sigma_i^2$ and $\hat{S}_t := \sum_{i=1}^t\hat{\sigma}_i^2$.

\subsection{Theoretical guarantees}

We now state the necessary assumptions to establish the primary inferential results.
\begin{assumption}
\label{ass:bounded_models}
\textup{(Bounded (Realized) Outcome Model Fitted Values)}.
There exists $N \in \mathbb{R}$ such that \[\underset{t\to\infty}{\text{lim sup}}|\hat{\mu}_w(X_t)| \leq N < \infty\] for all $w \in \mathcal{W}.$
\end{assumption}

Note that this assumption applies to the fitted values of the outcome models $\hat{\mu}_w$ evaluated on realized covariate vectors. In practice, this assumption is almost always satisfied: any reasonable outcome model will produce bounded predictions. In the rare event that fitted values exceed floating-point limits, researchers can safely cap them without losing meaningful information. Notably, our CSs do not depend on knowing $N$, and its value does not affect CS width.

Alongside Assumption~\ref{ass:bounded_models}, we also adopt the \textit{Bounded Realized Potential Outcomes} and \textit{At Least Linear Growth of Cumulative Conditional Variance} assumptions (Assumptions 1 and 2) of \citeauthor{liang_experimental_2024}. We now state the primary result:
\begin{theorem}
    \label{thm:thm_cs}
    Let $(\hat{\tau}_t)_{t=1}^\infty$ be the sequence of random variables where $W_t=w$ with probability $p_t^\text{MAD}(w)$, as in Definition~\ref{def:mad_probability}, with respect to some treatment assignment policy $\mathcal{A}$. Let \[\hat{V}_t = \sqrt{\frac{2(\hat{S}_t \eta^2 + 1)}{t^2 \eta^2}\text{\textup{log}}\left(\frac{\sqrt{\hat{S}_t \eta^2 + 1}}{\alpha}\right)}.\footnote{Note that $\eta$ is a tuning parameter that dictates the unit $t$ for which the CS is most precise. For all empirical examples, we follow the suggestions of \citeauthor{waudby-smith_time-uniform_2024} and \citeauthor{liang_experimental_2024} and define $\eta = \sqrt{\frac{-2\text{log}(\alpha) + \text{log}\left(-2\text{log}(\alpha) + 1\right)}{t^*}}$ where $t^*$ is the value of $t$ for which $\hat{V}_t$ is optimized (e.g. $t^* = 10,000$).}\]
    Under Assumptions 1 and 2 of Liang and Bojinov, and Assumption~\ref{ass:bounded_models}, $(\hat{\bar{\tau}}_t \pm \hat{V}_t)$ is a valid $(1-\alpha)$ asymptotic CS for $\bar{\tau}_t$ and $\hat{V}_t \overset{\text{a.s.}}{\to} 0$.
\end{theorem}

We defer the proof to Appendix~\ref{app:proof}. At a high level, the proof of Theorem~\ref{thm:thm_cs} follows by showing that $\delta_t = \omega(t^{-1/4})$---together with Assumptions 1 and 2 of \citeauthor{liang_experimental_2024} and our Assumption~\ref{ass:bounded_models}---ensures that $\hat{\bar{\tau}}_t$ satisfies a Lindeberg-type uniform integrability condition (see Lemma A.1 of \citeauthor{liang_experimental_2024}). This allows us to apply the universal asymptotic CS of \citet{waudby-smith_time-uniform_2024}.

In practice, Theorem~\ref{thm:thm_cs} lets researchers use MAD to combine the efficiency gains of adaptive experiments with valid, sequential ATE inference, while boosting inferential precision through covariate adjustment via flexible outcome models $\hat{\mu}_w(x)$.

\subsubsection{Additional considerations}

Finally, we will briefly touch on a number of relevant properties of MAD that hold for all our presented results. Since these properties are derived directly from MAD, we will in all cases refer the reader to the relevant discussion in \citeauthor{liang_experimental_2024}.
\paragraph{Non-stationarity} Theorem 2 of \citeauthor{liang_experimental_2024} guarantees that, under the conditions of Theorem~\ref{thm:thm_cs}, if $\bar{\tau}_t \to c$ for some $|c| > 0$ as $t \to \infty$, the sequence $(\hat{\bar{\tau}}_t \pm \hat{V}_t)$ will exclude 0 in finite time with probability 1. Intuitively, our CSs will allow the researcher to correctly reject the point null, $\mathcal{H}_0:\bar{\tau}_t=0$, in finite time. This property allows for non-stationarity as long as, in the long run, $\bar{\tau}_t$ converges to some value. For a thorough discussion of the implications of non-stationarity, particularly as it pertains to various stopping rules, see Section 5.1 of \citeauthor{liang_experimental_2024}.
\paragraph{Inference for multiple treatments and batched assignment settings}
As noted previously, all prior theoretical results extend directly to settings with $K \geq 2$ (multiple treatments) and to settings with batched assignment (Definition 5, Liang and Bojinov). To see a detailed formulation of these extensions, see Appendix~\ref{app:mad_covar_extensions}. The extension to settings with many treatments is crucial since, in practice, adaptive experiments are often deployed in settings with multiple treatments. As a result, we expect that most practitioners will be applying these methods in settings with $K \geq 2$.
\paragraph{Setting $\delta_t$} The deterministic sequence $\delta_t$ plays a critical role in the experimental design. As noted previously, $\delta_t$ determines the tradeoff between a fully random assignment policy and the adaptive assignment policy $\mathcal{A}$. As a result, \citeauthor{liang_experimental_2024} note that the regret of MAD (and thus our extensions) at any unit $t$ is a weighted sum of the regret of a fully random design and the regret of $\mathcal{A}$, where the weights are determined by $\delta_t$. Section 5 of \citeauthor{liang_experimental_2024} provides a detailed, finite-sample characterization of the regret of MAD, and Section 6 provides a robust set of recommendations for choosing $\delta_t$ to achieve various exploration/exploitation goals.

\subsection{Empirical results}

Using simulated data as well as data from a recent, large-scale randomized control trial (RCT), we demonstrate that MADCovar can achieve considerable precision gains relative to MAD. Additionally, these precision gains are robust to including (potentially many) irrelevant covariates in the outcome models.

\subsubsection{Simulations}

Using a grid of simulations, we evaluate MADCovar’s performance across increasing levels of covariate signal strength and number of irrelevant covariates included in our outcome models. This is motivated by the empirical concern that outcome–model–based estimators can suffer variance inflation when provided with many unrelated covariates.

\begin{figure}[!htbp]
\centering
\includegraphics[width = 0.8\textwidth]{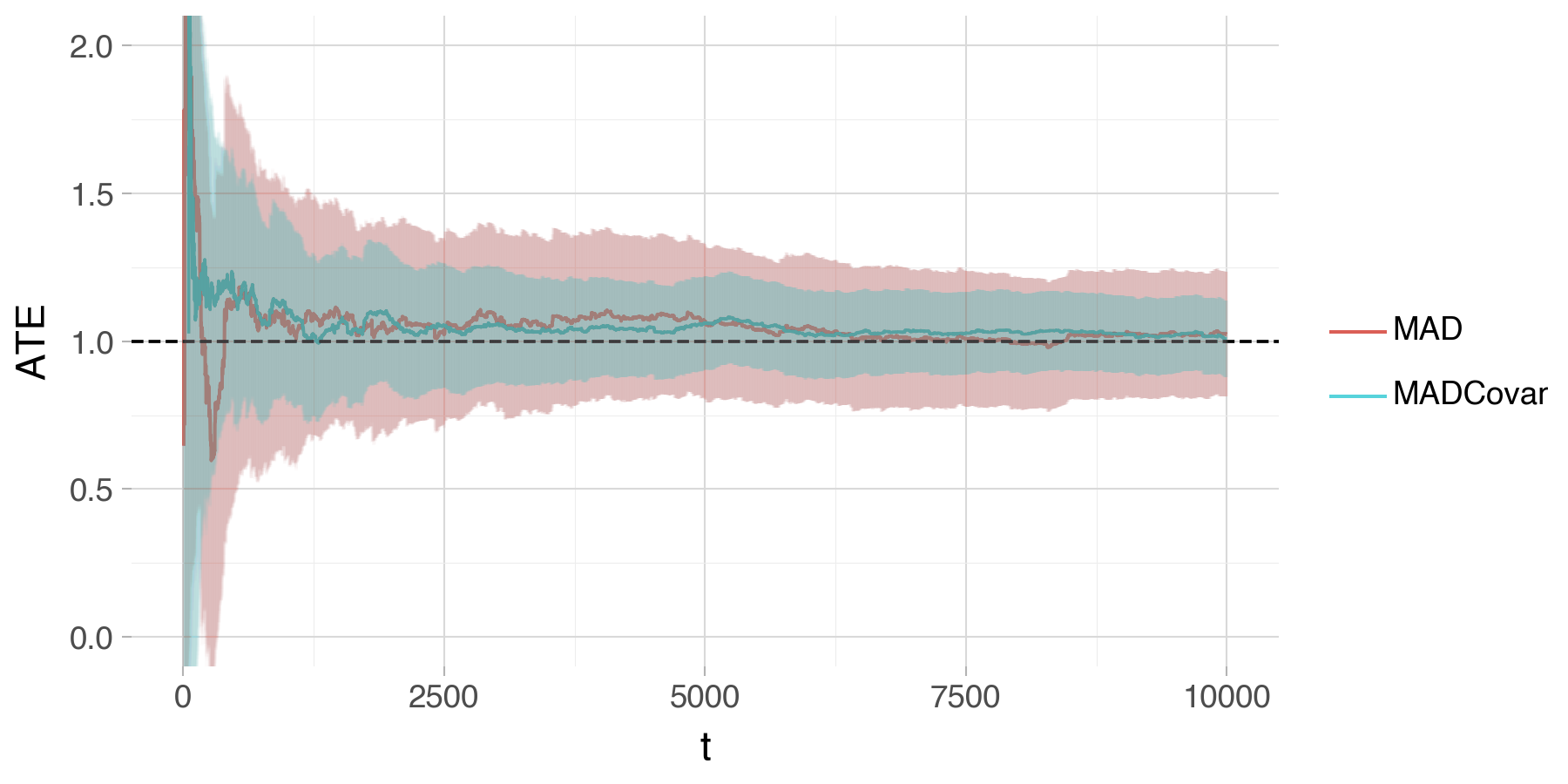}
\caption{MAD vs. MADCovar: comparison of 95\% confidence sequence widths across 10,000 units in a setting with moderate covariate signal.}
\label{fig:cs_precision_comparison}
\end{figure}

\begin{figure}[!htbp]
  \centering
  \includegraphics[width=0.9\textwidth]{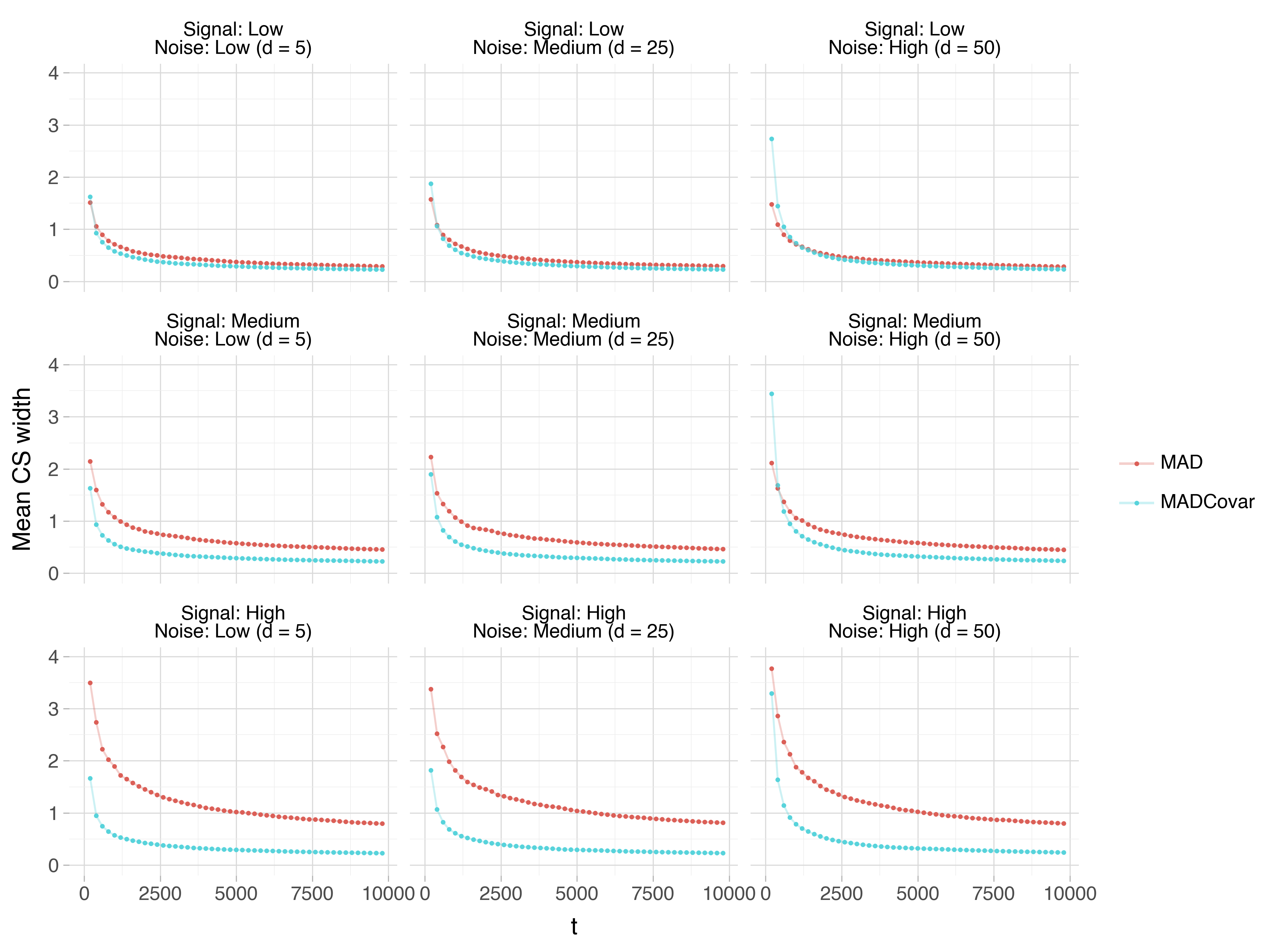}
  \caption{Comparing the mean CS width across 100 simulations for MAD vs. MADCovar over varying covariate signal strength values and number of irrelevant covariates.}
  \label{fig:mad_covar_robustness}
\end{figure}

We simulate a simple setting with a single treatment and control group; $w \in \{0, 1\}$.
Potential outcomes are defined as
\[
\begin{aligned}
Y_i(0) &= \gamma\,(0.3 X_{1,i} + 1 X_{2,i} - 0.5 X_{3,i}) + \varepsilon_i, \\
Y_i(1) &= 1 + \gamma\,(0.3 X_{1,i} + 1 X_{2,i} - 0.5 X_{3,i}) + \varepsilon_i.
\end{aligned}
\]
where
\[
X_{1,i},X_{2,i},X_{3,i} \sim \mathcal{N}(0,1) \text{ i.i.d.}, \quad \varepsilon_i \sim \mathcal{N}(0,1),
\]
and $\gamma \in [0, 1]$ controls covariate signal strength.
We vary \(\gamma\) at three levels---low (\(\gamma=0.1\)), medium (\(\gamma=0.5\)), and high (\(\gamma=1.0\))---to assess precision gains at various levels of covariate signal strength. We add irrelevant covariates \(X_{\text{irr},i}\sim\mathcal{N}(0,1)\) in three levels:
\begin{itemize}
  \item Low number of irrelevant covariates: 2 (total \(d=5\) covariates)
  \item Medium number of irrelevant covariates: 22 (total \(d=25\) covariates)
  \item High number of irrelevant covariates: 47 (total \(d=50\) covariates)
\end{itemize}
We set Thompson sampling as the underlying adaptive assignment algorithm (\(\mathcal{A}\)), a quickly-decaying sequence \(\delta_t = \frac{1}{t^{1/5}}\), and OLS for the outcome models (\(\hat{\mu}_w\)) fit on the full set of covariates $\{X_{1, i}, X_{2, i}, X_{3, i}, X_{\text{irr},i}\}$. Finally, we simulate MAD and MADCovar for 10,000 units and calculate the corresponding \(\hat{\bar{\tau}}_t\) estimates and 95\% CSs (see Figure~\ref{fig:cs_precision_comparison}). We repeat this process for 100 simulations and compare the mean CS interval width across these simulations for MAD and MADCovar (see Figure~\ref{fig:mad_covar_robustness}).

In settings with medium and high covariate signal, MADCovar provides precision gains between 40-70\% on average. Additionally, these gains remain consistent even when including many irrelevant covariates in the outcome model estimation. When utilizing many covariates (particularly unrelated covariates) in very small sample sizes ($N\leq1,000$), MADCovar may suffer from worse precision than MAD, however this is quickly rectified as the sample grows relative to the covariate space. This challenge can also be rectified by choosing appropriate outcome models, such as regularized regression models. 

\subsubsection{Large-scale RCT}

\begin{figure}[!htbp]
\centering
\includegraphics[width = 1\textwidth]{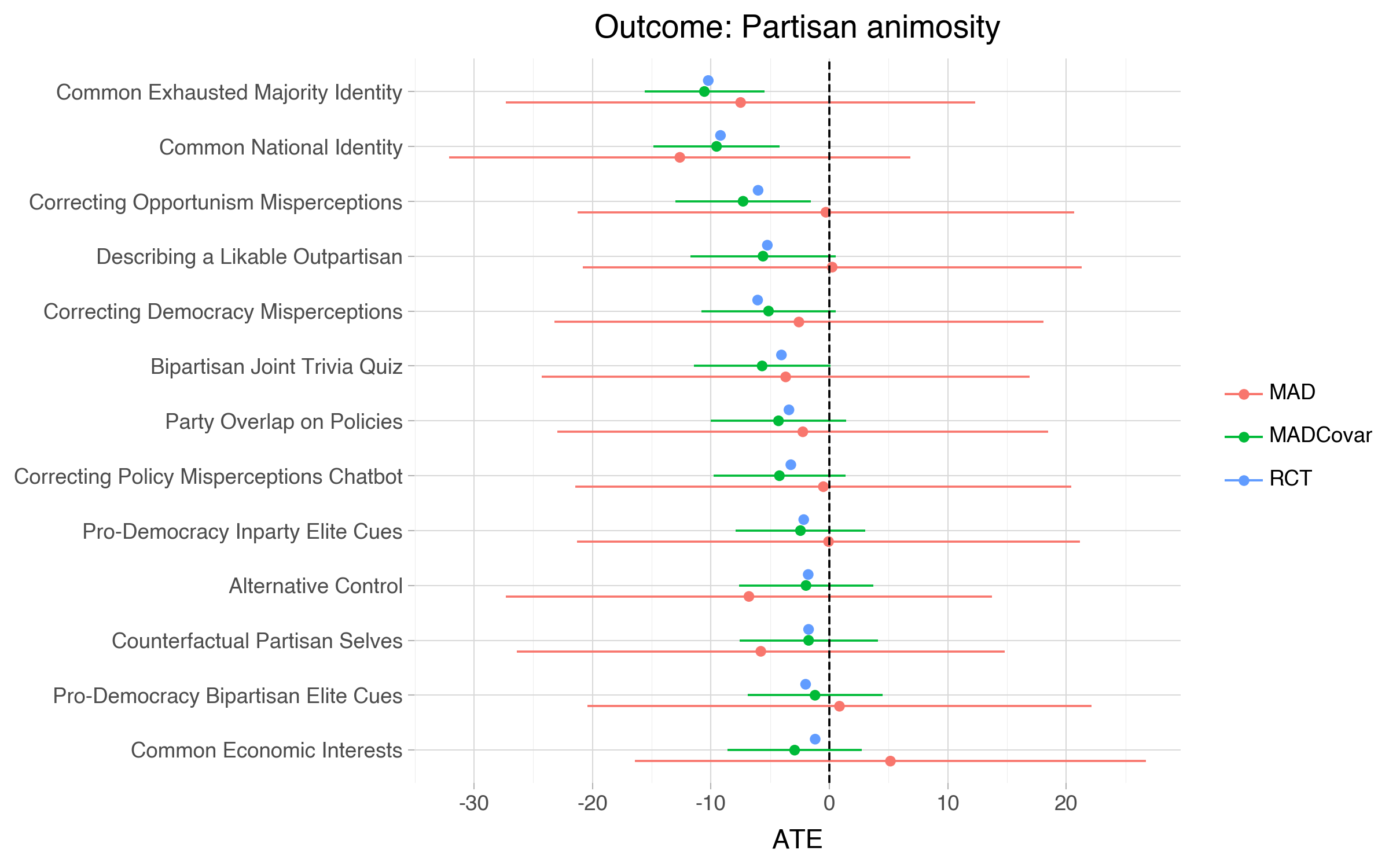}
\caption{Comparing ATE estimates and 95\% CS width for MADCovar vs. MAD relative to the ground truth (RCT ATE estimates) after running a simulated experiment with 32,000 participants.}
\label{fig:cs_rct_comparison}
\end{figure}

To further illustrate the substantial precision gains of MADCovar, we move beyond simulations and leverage large-scale RCT data from a mega-study by \citet{voelkel_megastudy_2024}. This study investigates the causal effects of various informational interventions on feelings of partisan animosity among Democrats and Republicans.
The study evaluates roughly 25 interventions using a fully randomized design, assigning approximately 1,000 respondents to each treatment and approximately 5,000 to the control group. The respondents are asked to record their feelings of partisan animosity on a numeric scale from 1-100 (higher meaning more animosity) both before and after the intervention. The study also collects a rich set of respondent characteristics that are posited to be predictive of the individual's level of partisan animosity. Using this dataset, we simulate an experiment by applying MAD and MADCovar to a random subset of 13 interventions, sampling units with replacement, to match the original RCT’s total sample size of $\approx 32,000$. As before, we set \(\mathcal{A}\) as Thompson sampling, \(\delta_t = \frac{1}{t^{1/10}}\), and OLS for the outcome models fit on the full set of respondent characteristics. At the conclusion of the experiment, we compare ATE estimates and 95\% CSs of MAD and MADCovar relative to the ground truth (RCT estimates). Figure~\ref{fig:cs_rct_comparison} demonstrates that the precision advantages of MADCovar extend to real-world scenarios, and are equally as striking as the precision gains observed in the simulations. In this example, covariate-adjusted ATE estimation allows us to achieve up to 60-75\% precision gains relative to the non-covariate-adjusted baseline.

\section{Power-modified Mixture Adaptive Design} 

Next, we introduce Power-modified MAD (MADMod) to address the challenge of imbalanced statistical power that is inherent in adaptive experiments. MADMod incorporates dynamic sample allocation to underpowered treatments, ensuring improved statistical power across all treatments. This allows researchers to balance the efficiency gains of adaptive experiments with the need for robust inference, enabling researchers to conduct experiments that are both statistically robust and resource efficient.

Let $K$ denote the set of treatment arms. For each arm $k \in K$ and unit $t$, we introduce importance weights $w_{k}(t) \in [0, 1]$ which are a function of $t$. Once the estimated ATE for arm $k$ reaches statistical significance, or equivalently the CS for arm $k$ excludes 0, $w_{k}(t)$ begins to shrink at a researcher-defined rate. By construction, the probabilities $\{p_t^\mathcal{A}(k)\}_{k \in K}$ form a valid probability distribution: $\sum_{k \in K}{p_t^\mathcal{A}(k)}=1$. We modify this distribution using a re-weighting function $g$, which adjusts $p_t^\mathcal{A}(k)$ based on its importance weight $w_{k}(t)$.

The re-weighted probability $g(p_t^\mathcal{A}(k)) := p_t^g(k)$ for each treatment arm $k$ at time $t$ is then computed as follows:
\begin{enumerate}
    \item \textbf{Generate Importance Weights}: For each treatment arm $k \in K$, set $w_{k}(t)=1$ if the arm has not yet reached statistical significance. Otherwise, set $w_{k}(t)=\text{decay}(t)$, where $\text{decay}(t)$ is a non-increasing function of $t$, such as polynomial decay.

    \item \textbf{Apply Importance Weights}: Each probability is then scaled by its importance weight: $p_t^*(k)=w_{k}(t)*p_t^\mathcal{A}(k).$

    \item \textbf{Compute Lost Probability Mass}: The probability mass lost due to down-weighting is: $L_t = \sum_{k \in K}{p_t^\mathcal{A}(k)*(1 - w_{k}(t))}.$

    \item \textbf{Compute Relative Redistribution Weights}: The total weight sum is: $W_t = \sum_{k \in K}{w_{k}(t)}.$ Each arm's share of the remaining mass is: $r_{tk} = \frac{w_{k}(t)}{W_t}.$

    \item \textbf{Redistribute Lost Mass}: Redistribute the lost mass proportionally to the relative weights: $p_t^g(k) = p_t^*(k) + (r_{tk} * L_t).$
\end{enumerate}
This transformation preserves the probability distribution $\big(\sum_{k \in K}{p_t^g(k)}=1\big)$ while allowing controlled reallocation of samples toward underpowered arms.

\subsection{Selecting the importance weight function}

While researchers can specify $w_k(t)$ as any non-increasing function of $t$, this function provides a flexible way to balance between two extremes: fully adhering to the underlying adaptive algorithm’s ($\mathcal{A}$) assignment probabilities or completely redirecting probability mass away from statistically significant arms. Setting $w_{k}(t) = 1$ retains the original adaptive probabilities, while $w_{k}(t) = 0$ eliminates allocation to statistically significant arms entirely. We discuss design choices that strike a middle ground between these two.

For researchers seeking modest power improvements while preserving the adaptive algorithm’s efficiency (i.e., favoring exploitation over exploration), a slowly decaying function such as $w_k(t) = \frac{1}{t^{1/10}}$ may be suitable. In contrast, if the primary goal is to ensure that as many treatments as possible are well-powered, a more rapidly decaying function like $w_k(t) = \frac{1}{t}$ is more appropriate. Alternatively, if the goal is to apply a fixed discount (e.g., 30\%) to the importance weights of statistically significant treatments, a function such as $w_k(t)=0.7$ achieves that. In general, users can select from a variety of decay functions to tailor this tradeoff to their specific goals.

\subsection{Empirical results}

Using simulations with $K=4$ treatment arms, we demonstrate MADMod’s ability to achieve precise, anytime-valid inference while improving statistical power across all treatments. Potential outcomes for the control arm are generated as i.i.d. draws from $Y_t(0) \sim \text{Bernoulli}(0.5)$, while treatment potential outcomes follow i.i.d. draws from $Y_t(k) \sim \text{Bernoulli}(p_k)$ for $k \in {1,2,3,4}$. We set $p=(0.6, 0.62, 0.8, 0.82)$ to create a clear separation between suboptimal arms (1 and 2) and near-optimal/optimal arms (3 and 4), corresponding to ATEs of 0.1, 0.12, 0.3, and 0.32. We set Thompson sampling as $\mathcal{A}$, and define $\delta_t = \frac{1}{t^{0.24}}$. Finally, we define a slowly-decaying importance weight function, $w_k(t)=\frac{1}{t^{1/8}}$, to achieve improved statistical power while still prioritizing the efficiency goals of the underlying Thompson sampling algorithm. We run 100 paired simulations comparing MAD and MADMod over 20,000 units with early stopping when all ATEs are detected as statistically significant.

\begin{figure}[!htbp]
\centering
\includegraphics[width = 1\textwidth]{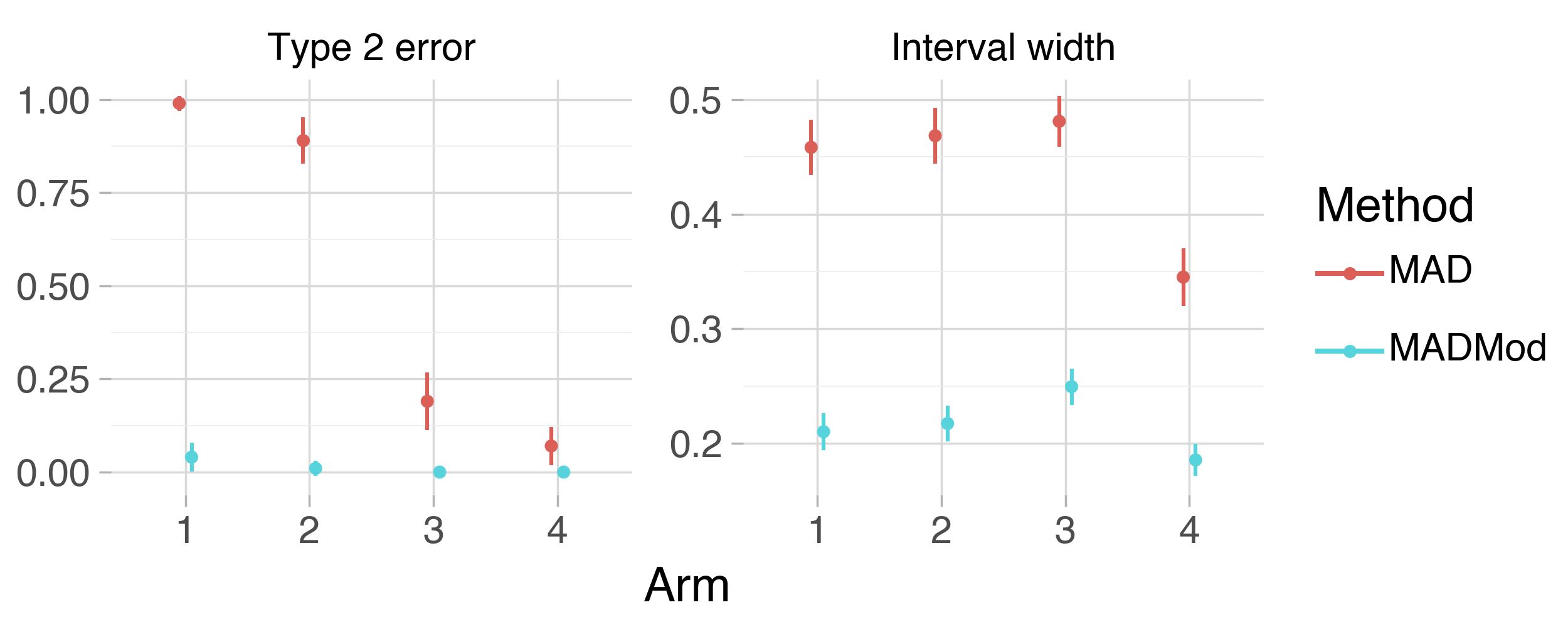}
\caption{Comparing the mean Type 2 error and confidence band width for the MAD vs. MADMod algorithms. Treatments 1 and 2 represent the sub-optimal treatment arms with harder-to-detect effect sizes. Treatments 3 and 4 represent the ``best'' treatment arms (i.e. the treatment arms that a MAB would choose to prioritize). The left panel displays mean Type II error and the right panel displays mean 95\% CS width.}
\label{fig:comparison_sim_error}
\end{figure}

Figure~\ref{fig:comparison_sim_error} shows that MADMod consistently improves Type 2 error control and precision across all treatment arms, with particularly large gains in error control for suboptimal treatments (treatment arms 1 and 2). Taken together, decreased Type 2 error and narrower interval widths allows for well powered and precise inference on all arms.

\begin{figure}[!htbp]
\centering
\includegraphics[width = 1\textwidth]{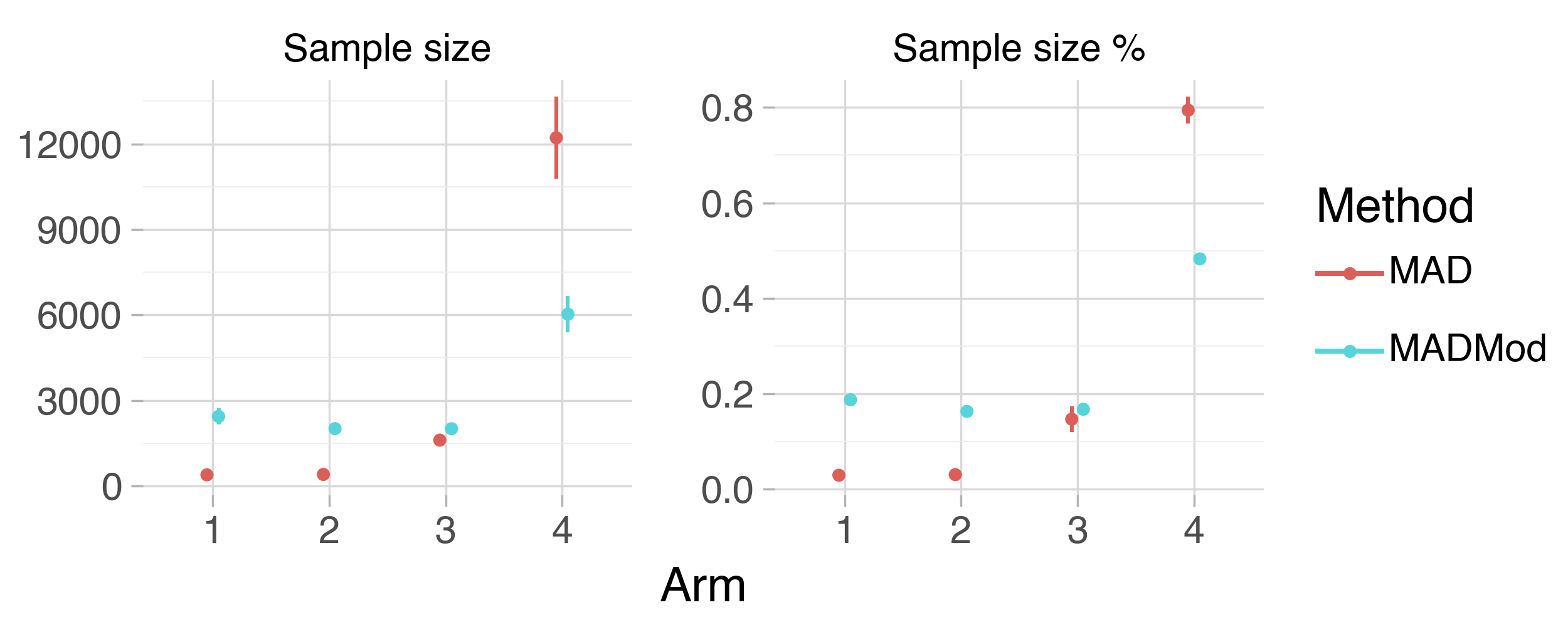}
\caption{Comparing the mean sample size and sample size percentage ($\pm 95\%$ confidence intervals) assigned to each treatment arm under the MAD vs. MADMod algorithms. Sample size is equal to the total number of participants within each treatment arm. Sample size \% is equal to the total number of participants assigned to a given treatment arm divided by the total number of participants in the entire experiment.}
\label{fig:comparison_sim_sample}
\end{figure}

As Figure~\ref{fig:comparison_sim_sample} shows, these gains come with a tradeoff in sample allocation. MADMod assigns significantly more samples to suboptimal arms than MAD, sacrificing some of the efficiency of $\mathcal{A}$ to ensure reliable inference for all arms. This tradeoff is often desirable, as researchers typically seek well-powered estimates for all treatment effects, not just the optimal arm.

\section{Limitations and broader impacts}

Our work offers methodological improvements that make anytime-valid inference in adaptive experiments more effective in applied settings. By increasing precision without requiring larger sample sizes, covariate-adjusted ATE estimation makes experiments more efficient and ethical, reducing costs, accelerating discovery, and limiting participant exposure to inferior treatments. These benefits are particularly important in high-stakes settings such as healthcare, education, and social policy. Moreover, improving statistical power across all treatments enables researchers to assess the effectiveness of every intervention---not just the most promising ones---reducing the risk of discarding treatments that, while not optimal, may still be valuable in practice.

Despite these clear benefits, our methods also face some limitations.

\subsection{MADCovar}

Anytime-valid inference is a powerful feature, enabling researchers to continuously monitor ATE estimates and stop data collection once a desired level of precision or statistical significance is reached. However, this flexibility comes with a trade-off in precision: for any fixed sample size $N$, the confidence sequence $\text{CS}_N$ will be wider, and thus less precise, than a corresponding confidence interval $\text{CI}_N$. In some cases, researchers may pre-specify $N$ in a non-data-driven way, whether due to ethical or fairness considerations, power calculations, or logistical constraints. In such settings, our CSs may offer less value, as they sacrifice precision for the anytime-valid property. Researchers may instead prefer fixed-sample ATE estimators that yield asymptotically valid CIs under adaptive sampling (e.g. \citet{hadad_confidence_2021}), offering greater precision when the stopping rule is fixed in advance.

\subsection{MADMod}

\citet{simchi-levi_multi-armed_2024} present a closely related approach that offers a rigorous theoretical treatment of the efficiency/power trade-off in adaptive experiments. They frame the problem as a minimax multi-objective optimization task and derive a Pareto-optimal solution, where improving either efficiency or statistical power necessarily compromises the other. In contrast, MADMod addresses the same trade-off through an algorithmic lens, allowing researchers to specify a function that governs how rapidly they trade efficiency for power.

While this flexibility comes at the cost of theoretical optimality, it offers important advantages. The method proposed by \citet{simchi-levi_multi-armed_2024}, though grounded in theory, is not broadly applicable across adaptive designs. Their solution is implemented via a new MAB algorithm tailored to their specific problem setting, which may not align with a researcher’s goals. MADMod, by contrast, inherits the versatility of MAD and can be integrated with virtually any adaptive experimental design. This allows researchers to retain control over the experimental strategy while still reaping the benefits of principled power reallocation.

\section{Summary}

In this paper, we advance methods for anytime-valid causal inference on the ATE in adaptive experiments, addressing key challenges for applied researchers. First, we introduce MADCovar, which leverages covariate adjustment to substantially improve ATE precision. Second, we present MADMod, which adaptively reallocates sample to under-powered treatment arms, enhancing statistical power across all arms. Together, these methods give practitioners greater flexibility and control over both inferential precision and statistical power in adaptive experimental settings.

\newpage

\bibliography{adaptive_inference}

\newpage

\appendix

\section{Proof of Theorem 1}
\label{app:proof}

The following proof is a direct copy of the proof of \citeauthor{liang_experimental_2024}'s (hereafter referred to as L\&B) Theorem 1 and is lightly adapted for our proposed ATE estimator $(\hat{\bar{\tau}}_t)$.

We begin by refreshing term definitions:
\begin{itemize}
    \item $\hat{S}_t := \sum_{i=1}^t\hat{\sigma}^2_i$
    \item $p_\text{min} := \text{min}(p_i^\text{MAD}(0), p_i^\text{MAD}(1))$
    \item $\delta_t$ is a decreasing, real-valued sequence $\delta_t \in (0, 1]$ s.t. $\delta_t = \omega(\frac{1}{t^{1/4}})$, where $\delta_t = \omega(\frac{1}{t^{1/4}})$ is equivalent to $1/\delta_t = o(t^{1/4})$
    \item $B_t := \sum_{i=1}^t \text{Var}[\hat{\tau}_i|\mathcal{F}_i]$
    \item $\eta$ is a hyper-parameter, the choice of which is discussed in Section 7 of L\&B
    \item $\alpha$ is a significance level; e.g. $\alpha = 0.05$
\end{itemize}
\textbf{Proof of Lemma A.1 (L\&B, Appendix 1)}
\begin{proof}
By Assumption 1 (L\&B) and our Assumption~\ref{ass:bounded_models}, 
\begin{align*}
    |\hat{\tau}_t| &= \left|\hat{\mu}_1(X_t) - \hat{\mu}_0(X_t) + \frac{\mathds{1}\{W_t=1\}(Y_t - \hat{\mu}_1(X_t))}{p_t^\text{MAD}(1)} - \frac{\mathds{1}\{W_t=0\}(Y_t - \hat{\mu}_0(X_t))}{p_t^\text{MAD}(0)}\right| \\
    &\leq 2N + \frac{M + N}{p_{\min}}.
\end{align*}
Next, since $|\tau_t| \leq 2M$, \[|\hat{\tau}_t - \tau_t| \leq 2(M+N) + \frac{M + N}{p_\text{min}},\] and thus
\begin{align*}
    (\hat{\tau}_t - \tau_t)^2 &\leq \left(2(M+N) + \frac{M + N}{p_{\min}}\right)^2 \\
    &\leq \frac{9(M + N)^2}{p_\text{min}^2}.
\end{align*}

Now, note that for all $w$, $p_t^\text{MAD}(w) \geq \delta_t(1/2)$ and so $\frac{1}{p_t^\text{MAD}(w)}=o(t^{1/4})$, and $(\hat{\tau}_t - \tau_t)^2 = o(t^{1/2})$ a.s..

Finally, as shown by L\&B via their Assumption 2, $B_t = \Omega(t)$. Therefore, $B_t^\kappa = \Omega(t^\kappa)$. Set $\kappa > 1/2$. Then, there exists some $\tilde{t}$ such that for all $t \geq \tilde{t}$, $\mathds{1}\{(\hat{\tau}_t - \tau_t)^2 > (B_t)^\kappa\}=0$, and so, \[\sum_{i=1}^t\frac{\mathrm{E}[(\hat{\tau}_t - \tau_t)^2\mathds{1}\{(\hat{\tau}_t - \tau_t)^2 > (B_t)^\kappa\}]}{(B_t)^\kappa} < \infty \text{ a.s.}\]
\end{proof}

\noindent\textbf{Proof of Theorem~\ref{thm:thm_cs}}

\begin{proof}
    Under Assumptions 1 and 2 (L\&B), and our Assumption 1, Lemma A.1 of L\&B holds. Together, Lemma A.1 and L\&B Assumption 2 satisfy Theorem 2.5 in \citet{waudby-smith_time-uniform_2024}. Let \[V_t^*:=\sqrt{\frac{2(B_t \eta^2 + 1)}{t^2\eta^2}\text{log}\left(\frac{\sqrt{B_t \eta^2 + 1}}{\alpha}\right)}.\] By Steps 1 and 2 of the proof of Theorem 2.5 in \citet{waudby-smith_time-uniform_2024}, $(\hat{\bar{\tau}}_t \pm V_t^*)$ is a valid $(1-\alpha)$ asymptotic CS. As noted in Section~\ref{sec:ate_estimation}, \[\text{Var}[\hat{\tau}_i | \mathcal{F}_i] \leq \sigma_i^2 \quad \text{where} \quad \sigma^2_i:=\frac{(Y_i(1)-\hat{\mu}_1(X_i))^2}{p_i^\text{MAD}(1)}+\frac{(Y_i(0)-\hat{\mu}_0(X_i))^2}{p_i^\text{MAD}(0)}.\] Hence, $(\hat{\bar{\tau}}_t \pm \tilde{V}_t)$ where \[\tilde{V}_t:=\sqrt{\frac{2(S_t \eta^2 + 1)}{t^2\eta^2}\text{log}\left(\frac{\sqrt{S_t \eta^2 + 1}}{\alpha}\right)}\] is still a valid $(1-\alpha)$ asymptotic CS, where $S_t=\sum_{i=1}^t\sigma_i^2$. This holds because replacing $B_t$ with $S_t$ only makes the CS wider since $S_t \geq B_t$.

    As noted in Section~\ref{sec:ate_estimation}, an unbiased estimator for $\sigma_i^2$ is \[\hat{\sigma}^2_i := \frac{\mathds{1}\{W_i=1\}(Y_i - \hat{\mu}_1(X_i))^2}{(p_i^\text{MAD}(1))^2} + \frac{\mathds{1}\{W_i=0\}(Y_i - \hat{\mu}_0(X_i))^2}{(p_i^\text{MAD}(0))^2}.\]

    To establish the validity of the CS in Theorem~\ref{thm:thm_cs}, we must show that $\frac{1}{t}\hat{S}_t - \frac{1}{t}S_t \overset{a.s.}{\to} 0$. Then, Condition L-3 of Theorem 2.5 of \citet{waudby-smith_time-uniform_2024} is satisfied and we can conclude that $(\hat{\bar{\tau}}_t \pm \hat{V}_t)$ where \[\hat{V}_t:=\sqrt{\frac{2(\hat{S}_t \eta^2 + 1)}{t^2\eta^2}\text{log}\left(\frac{\sqrt{\hat{S}_t \eta^2 + 1}}{\alpha}\right)}\] is still a valid $(1-\alpha)$ asymptotic CS. First, note that
    \begin{align*}
        (\hat{\sigma}_i^2)^2 &\leq \left(\frac{(M + N)^2}{p_i^\text{MAD}(1)^2} + \frac{(M + N)^2}{p_i^\text{MAD}(0)^2}\right)^2 \\
        &= \frac{(M + N)^4}{p_i^\text{MAD}(1)^4} + \frac{(M + N)^4}{p_i^\text{MAD}(0)^4} + 2\frac{(M + N)^2}{p_i^\text{MAD}(1)^4} \\
        &\leq \frac{(M + N)^4}{(\delta_i/2)^4} + \frac{(M + N)^4}{(\delta_i/2)^4} + 2\frac{(M + N)^2}{(\delta_i/2)^4} \\
        &\leq (M + N)^4\left(2^4\frac{1}{\delta_i^4} + 2^4\frac{1}{\delta_i^4} + 2^5\frac{1}{\delta_i^4}\right) \\
        &= o(i)
    \end{align*}
    where the last line follows because $\frac{1}{\delta_i}=o(i^{1/4})$. Define $X_i = \hat{\sigma}_i^2 - \sigma_i^2$ and note that $X_i$ is a martingale difference sequence. Hence,
    \begin{align*}
        \mathrm{E}[X_i^2] &= \mathrm{E}[(\hat{\sigma}_i^2)^2] - (\sigma_i^2)^2 \\
        &\leq \mathrm{E}[(\hat{\sigma}_i^2)^2] \\
        &= o(i).
    \end{align*}
    So, $\frac{\mathrm{E}[X_i^2]}{i^2} = \frac{o(i)}{i^2}$ and thus $\sum_{i=1}^\infty\frac{\mathrm{E}[X_i^2]}{i^2} < \infty$. For instance, if $\delta_i = \frac{1}{i^a}$ for some $0 \leq a < 1/4$, then there exists a constant $C < \infty$ and $i_0 > 0$ such that for all $i > i_0$, $\frac{\mathrm{E}[X_i^2]}{i^2} \leq Ci^{4a-2}$, and since $4a-2 < -1$, $\sum_{i=1}^\infty \frac{\mathrm{E}[X_i^2]}{i^2} \leq C \sum_{i=1}^\infty i^{4a-2} < \infty$ by the p-series test.

    Then, by the SLLN for martingale difference sequences (Theorem 1 of \citep{csorgo_strong_1968}), we can conclude that \[\frac{1}{t}\sum_{i=1}^tX_i \overset{a.s.}{\to} 0.\] Hence, Condition L-3 of \citet{waudby-smith_time-uniform_2024} is satisfied and by Step 3 of the proof for Theorem 2.5 of \citet{waudby-smith_time-uniform_2024}, we can conclude that $(\hat{\bar{\tau}}_t \pm \hat{V}_t)$ is still a valid $(1-\alpha)$ asymptotic CS.

    To show that $\hat{V}_t \overset{a.s.}{\to} 0$, note that
    \begin{align*}
        \hat{\sigma}_i^2 &\leq \frac{(M + N)^2}{p_i^\text{MAD}(1)^2} + \frac{(M + N)^2}{p_i^\text{MAD}(0)^2} \\
        &\leq \frac{(M + N)^2}{(\delta_i/2)^2} + \frac{(M + N)^2}{(\delta_i/2)^2} \\
        &= 8(M+N)^2\frac{1}{\delta_i^2}.
    \end{align*}
    Hence, $\hat{S}_t \leq 8(M+N)^2\sum_{i=1}^t \frac{1}{\delta_i^2} = 8(M+N)^2\sum_{i=1}^t o(i^{1/2})$ a.s.. Therefore, there exists positive real numbers $D$ and $x_0$ such that for all $t \geq x_0$, $\hat{S}_t \leq D\sum_{i=1}^t i^{1/2} < D\sum_{i=1}^t t^{1/2} = Dt^{3/2}$, allowing us to conclude that $\hat{S}_t = O(t^\frac{3}{2})$.

    Next, we show that $\text{log}\left(\hat{S}_t\right) = o(t^{1/2})$ a.s.. For $t \geq x_0$, \[\frac{\text{log}\left(\hat{S}_t\right)}{t^{1/2}} \leq \frac{\text{log}(Dt^{3/2})}{t^{1/2}} = \frac{\text{log}(D) + (3/2)\text{log}(t)}{t^{1/2}}\] and $\frac{\text{log}(D) + (3/2)\text{log}(t)}{t^{1/2}} \to 0$ as $t \to \infty$. Therefore, $\log\left(\hat{S}_t\right) = o(t^{1/2})$ a.s. and we conclude that \[\hat{S}_t\text{log}\left(\hat{S}_t\right) = o(t^2) \; \text{a.s.}\] Hence, $\hat{V}_t = \sqrt{\frac{2(\hat{S}_t \eta^2 + 1)}{t^2 \eta^2}\text{log}\left(\frac{\sqrt{\hat{S}_t \eta^2 + 1}}{\alpha}\right)} = o(1)$ a.s..
\end{proof}

As noted by L\&B, if $1/\delta_i$ was increasing at a rate faster than $i^{1/4}$ asymptotically, we are not guaranteed that $\hat{V}_t = o(1)$ a.s..

\section{MADCovar---extensions}
\label{app:mad_covar_extensions}

\subsection{Multiple treatments}

We formalize the results of Theorem~\ref{thm:thm_cs} for the setting with multiple treatment arms ($K\geq2$). We draw heavily from the exposition in Appendix C of \citeauthor{liang_experimental_2024}, and will refer the reader to this material for many of the details so as to avoid redundancy. 

We adopt the problem setting of the Generalized MAD (GMAD) laid out in \citeauthor{liang_experimental_2024}, Appendix C. In particular, $W_t \in \{0, 1, ..., K-1\}$ and GMAD has assignment probabilities $p_t^\text{MAD}(w) := \mathbb{P}_\text{MAD}(W_t=w|\mathcal{H}_{t-1})=\delta_t\left(\frac{1}{K}\right) + (1-\delta_t)p_t^\mathcal{A}(w)$.

For any pair of treatments $w, w' \in \{0, ..., K-1\}$, let $\tau_i(w, w')=Y_i(w)-Y_i(w')$ and define the ATE between $w$ and $w'$ up to unit $t$ as $\bar{\tau}_t:=\frac{1}{t}\sum_{i=1}^t\tau_i(w, w')$. So, under GMAD, our corresponding ATE estimator is \[\hat{\bar{\tau}}_t(w, w')=\frac{1}{t}\sum_{i=1}^t\hat{\tau}_i(w, w')\] where \[\hat{\tau}_i(w, w') := \hat{\mu}_w(X_i) - \hat{\mu}_{w'}(X_i) + \frac{\mathds{1}\{W_i=w\}(Y_i - \hat{\mu}_w(X_i))}{p_i^\text{MAD}(w)} - \frac{\mathds{1}\{W_i=w'\}(Y_i - \hat{\mu}_{w'}(X_i))}{p_i^\text{MAD}(w')}.\] We also have the corresponding upper bound on the variance: \[\mathrm{Var}[\hat{\tau}_i(w, w')|\mathcal{F}_i] \leq \sigma^2_i(w, w'), \text{ where } \sigma^2_i(w, w'):=\frac{(Y_i(w)-\hat{\mu}_w(X_i))^2}{p_i^\text{MAD}(w)}+\frac{(Y_i(w')-\hat{\mu}_{w'}(X_i))^2}{p_i^\text{MAD}(w')},\] and the analogous unbiased estimator of $\sigma_i^2(w, w')$: \[\hat{\sigma}^2_i(w, w') := \frac{\mathds{1}\{W_i=w\}(Y_i - \hat{\mu}_w(X_i))^2}{(p_i^\text{MAD}(w))^2} + \frac{\mathds{1}\{W_i=w'\}(Y_i - \hat{\mu}_{w'}(X_i))^2}{(p_i^\text{MAD}(w'))^2}.\] Finally, we let $S_t(w, w'):= \sum_{i=1}^t\sigma^2_i(w, w')$ and $\hat{S}_t(w, w'):=\sum_{i=1}^t\hat{\sigma}^2_i(w, w')$.

As in \citeauthor{liang_experimental_2024}, an analogous result of Theorem~\ref{thm:thm_cs} follows almost directly. As in the proof of Theorem~\ref{thm:thm_cs}, the only required modification to \citeauthor{liang_experimental_2024}'s proof is the additional constants introduced by our Assumption~\ref{ass:bounded_models} (Bounded Outcome Model Fitted Values). Otherwise, the remaining results and proof are identical; as such, we refer the reader to \citeauthor{liang_experimental_2024}, Appendix C, for a very thorough treatment of the details.

\subsection{Batched assignment setting}

We formalize the results of Theorem~\ref{thm:thm_cs} for the batched assignment setting, where the algorithm can only be updated at pre-defined points. Again, we draw heavily from the exposition in \citeauthor{liang_experimental_2024}, Appendix D, and will refer the reader to this material for many of the details.

We adopt the problem setting of the batched MAD, laid out in \citeauthor{liang_experimental_2024}, Appendix D. Let $\{(Y_i^{(j)}(w))_{w \in \mathcal{W}}\}$ be the set of all potential outcomes for unit $i$ in batch $j$. For any pair of treatments $w, w' \in \{0, ..., K-1\}$, we define the ATE \textit{within} a batch $j$ as: \[\tau_j^\text{batched}(w, w')=\frac{1}{B}\sum_{i=1}^B\left(Y_i^{(j)}(w) - Y_i^{(j)}(w')\right),\] and the corresponding unbiased estimator as \[\hat{\tau}_j^\text{batched}(w, w')=\frac{1}{B}\sum_{i=1}^B\hat{\tau}_i^{(j)}(w, w'),\] where \[\hat{\tau}_i^{(j)}(w, w') := \hat{\mu}_w^{(j)}(X_i) - \hat{\mu}_{w'}^{(j)}(X_i) + \frac{\mathds{1}\{W_i^{(j)}=w\}(Y_i^{(j)} - \hat{\mu}_w^{(j)}(X_i))}{p_j^{\text{MAD}_\text{batched}}(w)} - \frac{\mathds{1}\{W_i^{(j)}=w'\}(Y_i^{(j)} - \hat{\mu}_{w'}^{(j)}(X_i))}{p_j^{\text{MAD}_\text{batched}}(w')}\] and $\hat{\mu}_w^{(j)}$ is an arbitrary estimator of $\mathrm{E}[Y_i^{(j)}(w)|X=X_i,\mathcal{F}_{j}^\text{batched}]$. Then, for each $i=1, ..., B$, $\mathrm{Var}[\hat{\tau}_i^{(j)}(w, w')|\mathcal{F}_j^\text{batched}] \leq \sigma^{(j)2}_i(w, w')$, where \[\sigma^{(j)2}_i(w, w'):=\frac{\left(Y_i^{(j)}(w)-\hat{\mu}_w^{(j)}(X_i)\right)^2}{p_i^{\text{MAD}_\text{batched}}(w)}+\frac{\left(Y_i^{(j)}(w')-\hat{\mu}_{w'}^{(j)}(X_i)\right)^2}{p_i^{\text{MAD}_\text{batched}}(w')}.\] As a result, \[\hat{\sigma}_i^{(j)2}(w, w') := \frac{\mathds{1}\{W_i^{(j)}=w\}(Y_i^{(j)} - \hat{\mu}_w^{(j)}(X_i))^2}{(p_i^{\text{MAD}_\text{batched}}(w))^2} + \frac{\mathds{1}\{W_i^{(j)}=w'\}(Y_i^{(j)} - \hat{\mu}_{w'}^{(j)}(X_i))^2}{(p_i^{\text{MAD}_\text{batched}}(w'))^2}.\] Hence, \[\text{Var}[\hat{\tau}_j^\text{batched}(w, w')|\mathcal{F}_j^\text{batched}] = \frac{1}{B^2}\sum_{i=1}^B\text{Var}[\hat{\tau}_i^{(j)}(w, w')|\mathcal{F}_j^\text{batched}] \leq \frac{1}{B^2}\sum_{i=1}^B\sigma_i^{(j)2}(w, w').\] We define $S_b^\text{batched}(w, w') := \sum_{j=1}^b\frac{1}{B^2}\sum_{i=1}^B\sigma_i^{(j)2(w, w')}$ and $\hat{S}_b^\text{batched}(w, w') := \sum_{j=1}^b\frac{1}{B^2}\sum_{i=1}^B\hat{\sigma}_i^{(j)2}(w, w')$. Finally, we extend Assumption~\ref{ass:bounded_models} for the batched assignment setting.
\begin{assumption}{\textup{(Bounded (Realized) Outcome Model Fitted Values)}.} There exists $N \in \mathbb{R}$ such that \[\left|\hat{\mu}_w^{(j)}(X_t)\right| \leq N < \infty\] for all $j, i \in \mathbb{N}^+,w \in \mathcal{W}.$
\end{assumption}

As before, the remaining results and proof are identical to those in \citeauthor{liang_experimental_2024}; as such, we refer the reader to \citeauthor{liang_experimental_2024}, Appendix D, for a very thorough treatment of the details.

\section{Empirical confidence sequence coverage}

We empirically confirm the theoretical coverage guarantees provided by the $(1-\alpha)$ CSs for MADCovar ATE estimates. 

\subsection{Well-specified outcome model}

We modify the simulations in Section~\ref{sec:mad_covar} to have $K=6$ (five treatment arms), where potential outcomes are generated as i.i.d. draws from $Y_i(k) \sim \mathcal{N}(\mu_i(k), 1)$, where
\[\mu_i(k)
= 0.5
+ \beta_k
+ 0.3\,X_{1,i}
+ X_{2,i}
- 0.5\,X_{3,i},
\quad
k \in \{1,2,3,4,5\},
\]
and
\[
\beta_k =
\begin{cases}
0,   & k=0,\\
0.1, & k=1,\\
0.2, & k=2,\\
0.3, & k=3,\\
0.4, & k=4,\\
0.5, & k=5.
\end{cases}
\]
We set Thompson sampling as $\mathcal{A}$, define $\delta_t = \frac{1}{t^{0.24}}$, and estimate the following (well-specified) outcome model $Y_i = \alpha + \beta_1 X_{1,i} + \beta_2 X_{2,i} + \beta_3 X_{3,i} + \epsilon_i$ with OLS. We run MADCovar for 10,000 iterations and estimate ATEs and corresponding 95\% CSs. We repeat the simulation 1,000 times and calculate the empirical coverage error (see Table~\ref{tab:coverage_comparison_well_specified}), defined as the proportion of times the nominal 95\% CSs for the MADCovar ATE estimates failed to contain the true ATE, averaged over the simulation runs.

\subsection{Misspecified outcome model}

We now demonstrate the robustness of MADCovar's coverage guarantees to misspecification in the outcome model. We modify the data generating process in the previous simulation to involve non-linear terms and an interaction term:
\[\mu_i(k)
= 0.5
+ \beta_k
+ 0.3\,X_{1,i}^2
+ X_{2,i}*X_{3, i}
- 0.5\,\mathrm{e}^{X_{3,i}}.
\] We estimate the (now misspecified) outcome model $Y_i = \alpha + \beta_1 X_{1,i} + \beta_2 X_{2,i} + \beta_3 X_{3,i} + \epsilon_i$ with OLS. As before, we run 1,000 repeated simulations and calculate the empirical coverage error rate at the chosen significance level, $\alpha = 0.05$ (see Table~\ref{tab:coverage_scenario_misspecified}).

\begin{table}[ht]
\centering
\caption{Empirical coverage of 95\% CSs over 1,000 simulations.}
\label{tab:coverage_comparison}
\begin{subtable}[t]{\textwidth}
  \centering
  \caption{Well-specified outcome model}
  \label{tab:coverage_comparison_well_specified}
  \begin{tabular}{c | c}
    \toprule
    \textbf{Treatment arm} & \textbf{Coverage error ($\pm$ 95\% CI)} \\
    \midrule
    Arm 1 & 0.043 (0.03, 0.056) \\
    Arm 2 & 0.036 (0.024, 0.048) \\
    Arm 3 & 0.045 (0.032, 0.058) \\
    Arm 4 & 0.039 (0.027, 0.051) \\
    Arm 5 & 0.045 (0.032, 0.058) \\
    \bottomrule
  \end{tabular}
\end{subtable}

\vspace{1em}

\begin{subtable}[t]{\textwidth}
  \centering
  \caption{Misspecified outcome model}
  \label{tab:coverage_scenario_misspecified}
  \begin{tabular}{c | c}
    \toprule
    \textbf{Treatment arm} & \textbf{Coverage error ($\pm$ 95\% CI)} \\
    \midrule
    Arm 1 & 0.027 (0.017, 0.037) \\
    Arm 2 & 0.038 (0.026, 0.05) \\
    Arm 3 & 0.04 (0.028, 0.052) \\
    Arm 4 & 0.037 (0.025, 0.049) \\
    Arm 5 & 0.039 (0.027, 0.051) \\
    \bottomrule
  \end{tabular}
\end{subtable}
\end{table}

\end{document}